\documentclass[twocolumn,  prl,  aps,  superscriptaddress, longbibliography, showpacs, amsmath, amssymb, floatfix]{revtex4-1}   
\usepackage[normalem]{ulem} %\sout{}
\usepackage{float}
\usepackage{verbatim}
\usepackage{amsfonts}
\usepackage{bbm}
\usepackage{subfigure}
\usepackage{tipa}
\usepackage{color}
\usepackage{epstopdf}
\usepackage{epsfig}
\usepackage{subfigure}
\usepackage{mathrsfs} %include huati
\usepackage{amsmath, amssymb, amsfonts}
\usepackage{graphicx}% Include figure files
\usepackage{dcolumn}% Align table columns on decimal point
\usepackage{bm}% bold math
\usepackage{braket}
\usepackage{amssymb}% 罗马数字
\usepackage[colorlinks, linkcolor=blue, anchorcolor=blue, citecolor=blue, urlcolor=blue]{hyperref}
\makeatletter

\newcommand{\Rmnum}[1]{\expandafter\@slowromancap\romannumeral #1@}
\makeatother

\begin{document}
\title{Uncovering the origin of bound state in the continuum}
\author{Zeyu Rao}
\affiliation{Key Laboratory of Quantum Information,  University of Science and Technology of China,  Hefei 230026,  People’s Republic of China}
\affiliation{Anhui Province Key Laboratory of Quantum Network,  University of Science and Technology of China,  Hefei 230026,  China}
\affiliation{Hefei National Laboratory,  University of Science and Technology of China,  Hefei 230088,  China}
\author{Changling Zou}
\affiliation{Key Laboratory of Quantum Information,  University of Science and Technology of China,  Hefei 230026,  People’s Republic of China}
\affiliation{Anhui Province Key Laboratory of Quantum Network,  University of Science and Technology of China,  Hefei 230026,  China}
\affiliation{Hefei National Laboratory,  University of Science and Technology of China,  Hefei 230088,  China}
\affiliation{Synergetic Innovation Center of Quantum Information and Quantum Physics,  University of Science and Technology of China,  Hefei 230026,  China}
\author{Yang Chen}
\affiliation{Department of Precision Machinery and Precision Instrumentation,  University of Science and Technology of China,  Hefei 230026,  China}
\author{Guangcan Guo} 
\author{Ming Gong}
\email{gongm@ustc.edu.cn}
\affiliation{Key Laboratory of Quantum Information,  University of Science and Technology of China,  Hefei 230026,  People’s Republic of China}
\affiliation{Anhui Province Key Laboratory of Quantum Network,  University of Science and Technology of China,  Hefei 230026,  China}
\affiliation{Hefei National Laboratory,  University of Science and Technology of China,  Hefei 230088,  China}
\affiliation{Synergetic Innovation Center of Quantum Information and Quantum Physics,  University of Science and Technology of China, Hefei 230026,  China}

\date{\today}
	
\begin{abstract}
Bound state in the continuum (BIC) and quasi-BIC represent a remarkable class of wave functions that disobey conventional intuition by exhibiting spatially localized modes embedded in the continuum spectrum. In recent years, these states have found important applications in interdisciplinary systems as a non-radiating mode with ultra-long lifetime. In these applications, a key question is how to convert a quasi-BIC into an exact BIC, and what the general criterion is for this transition. In this work, we uncover its origin using two steps in a two-band model with a arbitrary confining potential. Firstly, we demonstrate that a bound state coupled to a continuum band can yield quasi-BIC. Then, we show that tuning the coupling between the bands can convert the quasi-BIC into an exact BIC. In our theory, the real and complex poles of the spectra have a clear physical meaning for the quasi- and exact BICs, and we give the general criterion for exact BICs. Unlike previous proposals, our theory requires neither symmetry protection nor topological constraints and can be extended to a multiband model, providing a new framework for realizing BICs and offering new insights for their design in different fields, including photonics, acoustics, ultracold atoms and Bose-Einstein condensate with and without many-body interactions. 
\end{abstract}
\maketitle

Conventional bound states are generally realized using confining potentials, yielding discrete energy levels with prolonged lifetimes. However, it is still possible to construct a bound state in the continuum (BIC), which, just as the name implies, will embed in the energy range of the extended states. This idea can be traced back to von Neumann and Wigner \cite{von1929uber, Stillinger1975BoundStates, hsu2013observation}, which can be constructed using $E = (T\psi)/\psi - V(r)$, where $T = p^2/2m$ is the kinetic energy operator, $m$ is the mass, $V(r)$ is the potential, and $\psi$ is a square integrable wave function. When $\psi$ and $V(r)$ are properly chosen, $E$ can lie in the continuum band for BIC, which is a dark mode with an infinitely long lifetime. In practice, however, this state is rather fragile and any perturbation or imperfection in the potential $V(r)$ can lead to coupling between BIC and the continuum state, yielding quasi-BIC with a long tail and with finite lifetime \cite{hsu2013observation}. The BIC and quasi-BIC achieve their confinement through some delicate interference effects \cite{Friedrich1985interfering, Shanhui2003Temporal,weimann2013compact,rybin2017high,Azzam2018formation}, symmetry protections \cite{Parker1966Resonance, Parker1967Resonance, Cobelli2009experimental, Moiseyev2009Suppression, Cederbaum2003Conical, Plotnik2011Experimental, Dreisow2009Adiabatic,Ni2016Tunable} or topological mechanisms \cite{zhen2014topological, yang2013topological,Helical2015Vladimir}, yielding non-radiating mode even embedded in a radiative continuum field. This mode has profound implications in photonics \cite{kang2023applications, Lee2012observation, Marinica2008Bound, koshelev2023bound, zhou2023increasing, minkov2019doubly, Yang2014Analytical, Vaidya2021Point, Optically2014Boretz,Bulgakov2014Bloch,yoon2015critical,Ni2016Tunable},  metasurfaces \cite{zhang2022chiral, Gorkunov2020Metasurfaces, santiago2022resonant, Dixon2021Self, Koshelev2018Asymmetric, shi2022planar,tittl2018imaging,leitis2019angle} and condensed matter systems \cite{Guessi2015Quantum, Guevara2003Ghost, González2010Bound,bound2006Sadreev,Helical2015Vladimir}. These states have attracted significant attention due to their ultrahigh quality factors \cite{Lee2012observation, hsu2013observation} and strong light-matter interactions \cite{Lee2012observation, zhou2023increasing, Ha2018Directional},  offering unprecedented opportunities for nonlinear optics \cite{Carletti2018Giant, Liu2019High, koshelev2020subwavelength, wang2020doubly, santiago2022resonant},  lasing \cite{yu2021ultra, kodigala2017lasing, hwang2021ultralow, zhang2022chiral, sang2022topological,huang2020ultrafast}, sensing \cite{meudt2020hybrid, leitis2019angle, liu2017optical, tseng2020dielectric, altug2022advances, yesilkoy2019ultrasensitive, jahani2021imaging, leitis2021wafer} and quantum information processing \cite{Zheng2013Persistent, van2013photon}. Moreover, it has attracted much attention in acoustics \cite{Deriy2022Bound,kronowetter2023realistic,Jia2023Bound,marti2024observation}, ultracold atoms \cite{kartashov2017boundstates, Huang2024Interaction}, quantum transport \cite{MARTINEZ2025Uncovering, Pinto2024Bound, Moiseyev2009Suppression, Kim1999Resonant}, exciton polariton \cite{wu2024exciton} and many other fields \cite{Valero2025Exceptional,Xiao2017Topological,Schiller2024Time,An2024Multibranch,chen2016mechanical,zou2015guiding}. 

In these researches, a key issue is to understand the origin of converting a quasi-BIC into an exact BIC. We are interested in the bound state that is square integrable in the whole space \cite{bicperiodic}. Here, we uncover its origin using a simplified two-band model with an arbitrary confining potential. We first demonstrate the quasi-BIC by coupling a localized state using an attractive $\delta(x)$ potential to a continuum band, and then we consider a generalized confining potential and give the general criterion for converting the quasi-BIC to an exact BIC. We give the results for the exact BIC using two different models. In this theory,  we show that the position of the real and complex poles, having a clear physical meaning of localization, is essential for the formation of BIC, and the mechanism can be applied in the multiband model. We hope this new approach can provide a general guideline to realize the BIC in realistic experiments, yielding intriguing applications in photonic and acoustic structures, ultracold atoms, and Bose-Einstein condensates with or without many-body interactions. 

\begin{figure}[htbp]
	\centering
 \includegraphics[width=0.48\textwidth]{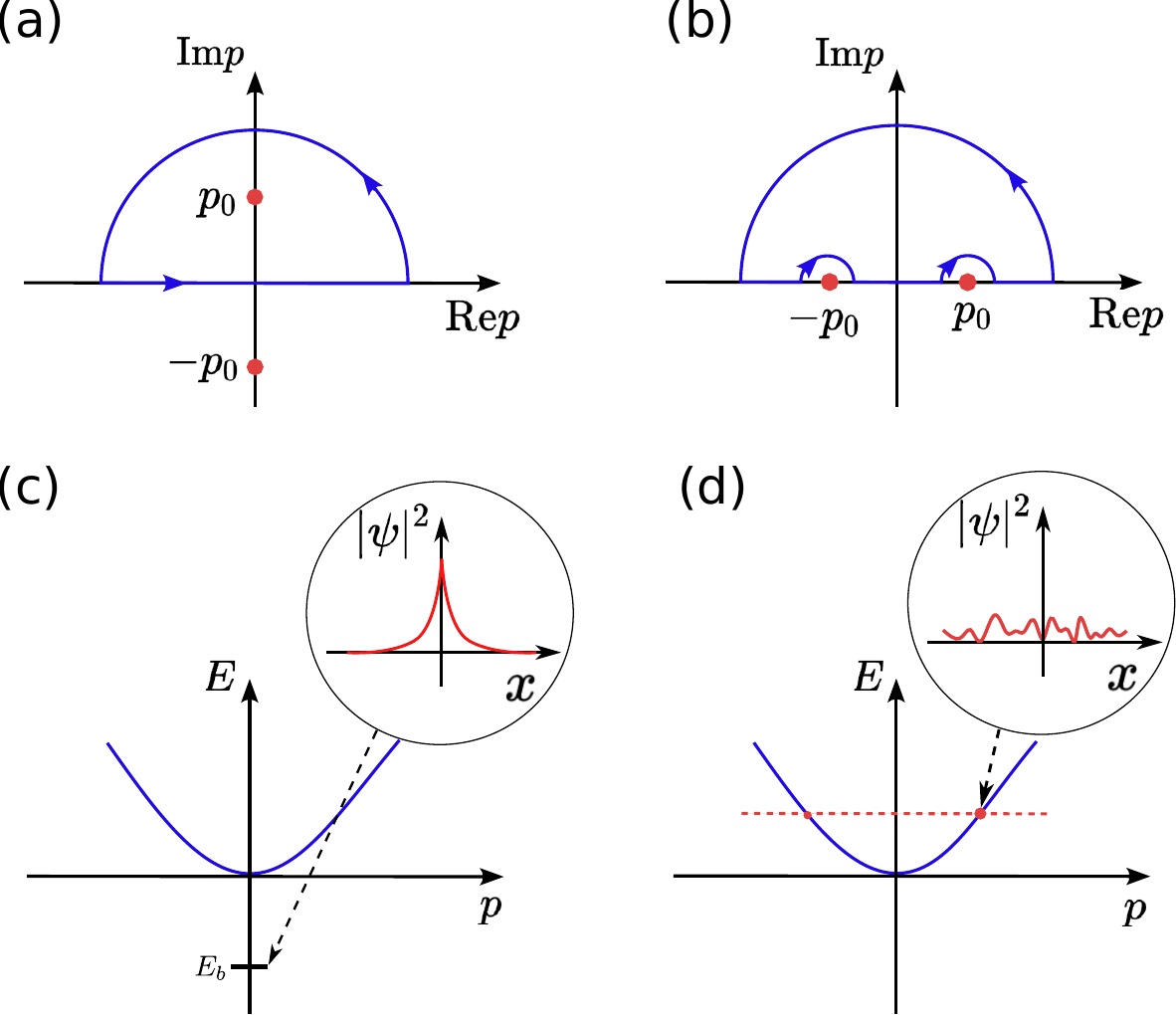}
\caption{(a) The contour integral when calculating the wave function from $\psi_p$. The red point means the complex poles of $\psi_p$,  contributing an exact bound state. (b) Real poles along the real axis for an extended (or non-normalizable) wave function. (c) The dispersion relation of the $\delta$ function potential system. The energy of the bound state ($E_{\text{b}}$) and the associated wave function are shown. (d) The energy of the extended case. The red point indicates the momentum (or pole) of the wave function. }
\label{fig-fig1}
\end{figure}

{\it (I) Bound state with $\delta$ function potential}: Our basic idea is related to interference  resonance for BIC  by Friedrich and Wintgen \cite{Friedrich1985interfering}, which studies the coupling between two closed channels (bound states) with one open channel (continuum state) via Feshbach resonance. %\cite{Moiseyev2009Suppression}.
However, our approach is much more transparent, general, and can be extended to arbitrary potential $V(x)$ with $N$ bands, thus the transition from a quasi-BIC to an exact BIC can be seen much clearly. We warm up by considering the one-dimensional $\delta(x)$ potential with Hamiltonian 
\begin{equation}
H = {p^{2} \over 2m} + \lambda \delta(x), 
\end{equation}
where $p=-i\partial_{x}$,  setting $\hbar=1$. The wave function can be written as  $\psi(x)=\sum_p \frac{\lambda e^{ipx}}{E-p^{2}/(2m)} \psi(0)$. It is well-known that when $\lambda < 0$,  there will be a solution when $E < 0$,  with $\psi(0) \ne 0$. Then integration of $p$ using the contour integral in the upper half-plane in Fig. \ref{fig-fig1} (a),  with complex momentum poles at $p = \pm p_{0} = \pm \sqrt{2mE} = \pm \sqrt{2m|E|} i$, will yield the bound state energy 
\begin{equation}
1 = {2m\lambda \over 2\pi} {2\pi i \over 2\sqrt{2m |E_\text{b}|}i},  \quad E_\text{b} = -{\lambda^2 m \over 2}.
\label{eq-EbinmodelI}
\end{equation}
The wave function of the localized state will become $\psi(x) \sim \exp(-\sqrt{2m|E|} |x|)$, see Fig. \ref{fig-fig1} (c). Furthermore,  $E_b$ can be determined from the boundary condition
\begin{equation}
-{1\over 2m} (\psi'(0^+) - \psi'(0^-)) + \lambda \psi(0) = 0.
\label{eq-continuumcondition}
\end{equation}
When $\lambda > 0$,  the bound state will not exist,  yielding $\psi(0) = 0$ in the upper equation. In this case, the poles are real, and for $x > 0$,  we can calculate the integral along the loop in Fig. \ref{fig-fig1} (b). And the integral gives 
\begin{equation}
{1\over 2\pi} \int dp {e^{ipx} \over E - p^2/(2m)} = \frac{m}{p_{0
}} \sin(p_0x)\text{sign}(x), 
\label{eq-contour-1d}
\end{equation} 
yielding the extended state in Fig. \ref{fig-fig1} (d),  with zero density at the singular potential. One may check immediately that this state is the eigenvector of $H$. Here, the heuristic discussion of this model aims to show that the position of the poles is essential for the bound states. 

\begin{figure}[htbp]
	\centering
 \includegraphics[width=0.48\textwidth]{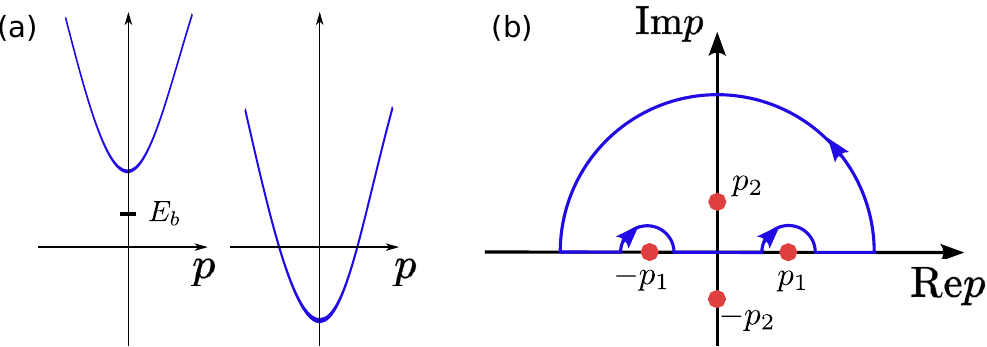}
 \caption{(a) General approach for quasi-BIC using a $\delta$ function potential and two bands,  in which the bound state with energy $E_b$ (left) is embedded in the continuum band (right). (b) Contour integral with four poles,  two of which are real and the others are complex. }
 \label{fig-fig2}
 \end{figure}

{\it (II) Two-band model with a single $\delta$ function potential and quasi-BIC}: We then ask the general question of what the fate of the above bound state interacting with a continuum band. If the coupling is weak enough,  the eigenvalue of this mode should not be changed dramatically;  thus, it is quite possible that this state can still partially survive. However, this mode interacts with the extended bands, presenting a possible oscillating wave in the long-distance limit,  yielding a quasi-BIC. Thus, strictly speaking, this mode should be extended. We address the above issue using the following model 
\begin{equation}
H = {p^2 \over 2m} + A+ \delta (x) B,  \quad B = \begin{pmatrix}
    \lambda  & 0 \\ 
    0 & 0  
\end{pmatrix},  \quad A = \mu \sigma_z + g\sigma_x,  
\end{equation}
where $\sigma_z$ and $\sigma_x$ are Pauli matrices. This model is also schematically presented in Fig. \ref{fig-fig2} (a). The second term indicates that only one of the bands has a $\delta$ function potential for the bound state, and the other band plays the role of a continuum spectrum. The term $\mu$ is used to shift the energy difference between the two bands,  and $g$ is the direct coupling between the two bands. If $g = 0$,  the two bands are decoupled,  yielding a bound state in one of the bands. For convenience,  we set
\begin{equation}
    H_{0}(p)={p^2 \over 2m} + A.
\end{equation}
Using the Fourier transformation,  similar to that in case (I),  will yield the following representation in momentum space via $\psi(x)=\frac{1}{2\pi}\int \psi_p e^{ipx}dp$, we have 
\begin{equation}
    \psi_p=(E-H_{0}(p))^{-1} B \psi(0), 
\label{eq-psi_p}
\end{equation}
where $\psi(0)$ is the amplitude of wave function $\psi(x=0)$. It leads to  $
    (E-H_{0}(p))^{-1}= (g\sigma_{x}+E-\frac{p^{2}}{2m}+\mu\sigma_{z})/\text{det}(E-H_0(p))$,   where $\text{det}(E-H_0(p))=E^{2}-g^{2}+\frac{p^{4}}{4m^{2}}-\frac{Ep^{2}}{m}-\mu^{2}$.

The integration of momentum $p$ in a closed loop will show that the form of the wave function is only determined by the poles of $\psi_p$. The roots of $f(p)=0$ are $p=\pm\sqrt{2m(E\pm\sqrt{g^{2}+\mu^{2}})}$. As a result, when $E > \sqrt{g^2 + \mu^2}$,  all the four poles are real (in the real axis); and when $E < - \sqrt{g^2 + \mu^2}$,  all the four poles are pure complex (in the imaginary axis). Between these two conditions,  we have two real and two complex poles, and the third condition is most promising for BIC for the mixing of extended and bound states.  Let us define  $p_{1}=\sqrt{2m(E+\sqrt{g^{2}+\mu^{2}})}$,  $p_{2}=\sqrt{2m(E-\sqrt{g^{2}+\mu^{2}})}$,  and $\pm p_{1, 2}$ are associated with the four poles.  We are interested in the condition that $-\sqrt{g^{2}+\mu^{2}}<E<\sqrt{g^{2}+\mu^{2}}$,  where $p_{1}$ is real,  and $p_{2}$ is purely imaginary,  generating the mixing of localized and extended function $\psi(x) = (\psi_1(x),  \psi_2(x))^T$ with 
\begin{equation}
\psi = a \psi_1 \exp(-|p_2||x|) + b \psi_2 \sin(p_1x) \text{sign}(x), 
\label{eq-wavefunction-modelII}
\end{equation}
where $\psi_1 = (\mu + \sqrt{g^2 +\mu^2},  g)^T$ and $\psi_2 = (\mu - \sqrt{g^2 +\mu^2},  g)$,  and $a/b=p_{1}/\vert p_{2}\vert$.

We next check that the wave function in Eq. \ref{eq-wavefunction-modelII} is the eigenvector of $H$. When $x \ne 0$,  this is obvious for any $E$. Thus, we have to verify its boundary condition that 
\begin{equation}
-{1\over 2m} (\psi'(0^+) - \psi'(0^-)) + B \psi(0) = 0, 
\label{eq-boundaryconditionB}
\end{equation}
yielding $p_2 = m\lambda/2 (1 + \mu/\sqrt{g^2 + \mu^2})$,  and $E_b = \sqrt{g^2 + \mu} -p_2^2/(2m) $,  which gives bound state energy
\begin{equation}
    E_{\text{b}}= -{\lambda^{2}m \over 8}  (1 + \frac{\mu}{\sqrt{g^{2}+\mu^{2}}})^2+\sqrt{g^{2}+\mu^{2}}.
    \label{eq-boundenergyB}
\end{equation}
It should satisfy $E_{\text{b}}\in(-\sqrt{g^{2}+\mu^{2}}, \sqrt{g^{2}+\mu^{2}})$.  Obviously,  when $g= 0$, $E_\text{b} = \mu - \lambda^2m/2$, yielding Eq. \ref{eq-EbinmodelI}. Thus $\lambda$ should be in the range of $\lambda_c < \lambda < 0$,  with
\begin{equation}
    \lambda_c = -{4(\sqrt{g^{2}+\mu^{2}})^{3/2} \over \sqrt{m}(\sqrt{g^{2}+\mu^{2}}+ \mu)}.
\end{equation}
When $\lambda < \lambda_c$,  the quasi-BIC will become an exact bound state. It should be emphasized that the extended mode $\psi_e \propto \psi_2 \sin(p_1x) \text{sign}(x)$ is not the eigenvector of the Hamiltonian $H$, since this wave function can not satisfy the continuum condition in Eq. \ref{eq-boundaryconditionB}. We show that this is a typical feature of the 
$\delta$ function potential with the stringent boundary condition in Eq. \ref{eq-continuumcondition}. For this reason,  we can only realize quasi-BIC. 

{\it (III) Two-band model with $\delta$ function potential,  BIC and quasi-BIC}: With the above result, we next explore the possible fate of this state using the more generalized model
\begin{equation}
H = {p^2 \over 2m} + A + \delta (x) B,  \quad B = \begin{pmatrix} b_{1} & b_{2} \\ b_{2} & b_{3} \end{pmatrix}, 
\end{equation}
where $B$ is a Hermite matrix. The discussion of this part is quite similar to case (II).  We will find that the wave function can be made up of two parts:  an exponentially localized state and an extended state,  both of which are even functions of $x$.  It can be written in the same form as Eq. \ref{eq-wavefunction-modelII}, in which  the coefficients of $a$ and $b$ can be determined using the contour integral method,  yielding complex momentum poles at $\pm p_2$ and real poles $\pm p_1$ for the localized modes and extended modes, respectively. The bound state energy can be determined by the boundary condition of Eq. \ref{eq-boundaryconditionB},  yielding $a/b = p_1 (\sqrt{g^2 + \mu^2} - \mu)/(b_2 gm + (b_1m - |p_2|)\sqrt{g^2 + \mu^2})$. Thus, the $B$ matrix will influence the ratio between these two wave functions. Furthermore, we can obtain 
\begin{equation}
|p_2| = {m (2b_2g + b_3 (\sqrt{g^2 + \mu^2} - \mu ) + b_1 (\sqrt{g^2 + \mu^2} + \mu)) \over 2\sqrt{g^2 + \mu^2}}, 
\end{equation}
yielding binding energy $E_\text{b} =  \sqrt{g^2 + \mu^2} -|p_2|^2/(2m)$. Obviously,  this state always exists even when some of the $b_i$ are positive for repulsive interaction,  when $|E_\text{b}| < \sqrt{g^2 + \mu^2}$. Let us denote the extended modes as $\psi_e = b \psi_2 \sin(p_1x) \text{sign}(x)$, and we can check immediately that this state alone can not satisfy the above boundary condition. Thus, for a general $B$ matrix,  we can only obtain quasi-BIC; yet the form of this state can be controlled with more flexibility using this $B$ matrix. 

{\it (IV) Generalization to confining potentials}:  The above models are stimulating. It presents a much more straightforward approach to realizing the quasi-BIC. However,  the requirement of the continuum condition by the $\delta$ function potential prohibits the extended state along the eigenvector of the Hamiltonian, yielding only quasi-BIC,  but not exact BIC. This $\delta(x)$ potential is hard to realize in experiments; yet it can be approximated very well using a confining potential, indicating possible generalization. Next, we aim to derive the general criterion for the existence of BIC in a confining potential in the two-band models,  showing that the above mechanism, in which the BIC is related to the poles, is applicable to much wider physical systems.

To this end,  let's consider the following general model
\begin{equation}
H = {p^2 \over 2m} + A + V(x) B, 
\end{equation}
where $B$ is a $2\times 2$ Hermite matrix and $V(x)$ is a confining potential. As discussed before, in order to obtain the quasi- or exact BIC, we assume the system energy lies between the two bands. We consider two different cases. Firstly,  we maintain the definition of $A$ as before, that is, a constant matrix, independent of momentum. Using the same approach 
\begin{equation}
\begin{aligned}
    \psi(x)&=B_{1}\int_{-\infty}^{\infty} \exp(-\vert x-x'\vert\vert p_{2}\vert)V(x')\psi(x')dx' \\
    &+B_{2}\int_{-\infty}^{\infty} \sin{(\vert x-x'\vert p_{1})} V(x')\psi(x')dx', 
    \label{eq-psi_x}
\end{aligned}
\end{equation}
where $B_{1}=a\begin{pmatrix}
    \psi_{1} & c_{0}\psi_{1}
\end{pmatrix}B$,  $B_{2}=b\begin{pmatrix}
    \psi_{2} & d_{0}\psi_{2}
\end{pmatrix}B$,  $c_{0}=(-\mu+\sqrt{g^{2}+\mu^{2}})/g$,  $d_{0}=(-\mu-\sqrt{g^{2}+\mu^{2}})/g$,  $a=-m/(2\sqrt{g^{2}+\mu^{2}}\vert p_{2}\vert)$,  $b=-m/(2\sqrt{g^{2}+\mu^{2}}p_{1})$. The above equation gives a quasi-BIC solution. Now we explore what makes this quasi-BIC solution an exact BIC.

\begin{figure}[H]
	\centering
 \includegraphics[width=0.48\textwidth]{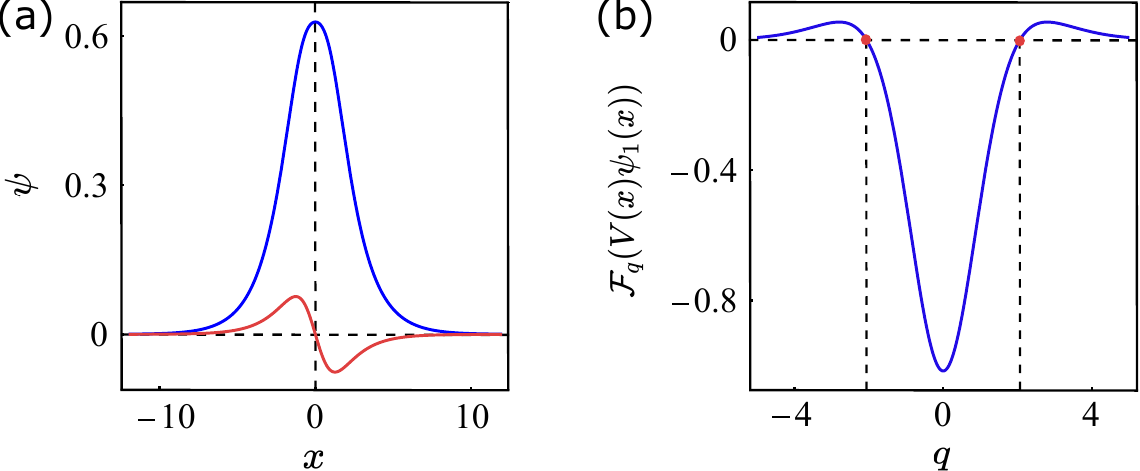}
 \caption{(a) The profile of BIC $\psi(x)=(\psi_{1}(x), \psi_{2}(x))^{T}$ with $\psi_{1}$ (blue) and $\psi_{2}$ (red),  with parameters $\gamma=0.5$,  $ \nu=0.7$,  $\mu =1$. (b) Fourier transformation of $V(x')\psi_1(x')$ in momentum space,  in which the constraint Eq. (\ref{eq-BFqcomponent2}) is satisfied,  with real poles at $q = \pm 2.06325. $} 
 \label{fig-fig3}
 \end{figure}

The second key idea is the following. To convert the quasi-BIC into an exact BIC,  some new constraints are required. We need the long oscillating tail included in the sine term to vanish. Let us assume 
\begin{equation}
V(x')\psi(x') = \sum_q \mathcal{F}_q\left(V(x')\psi(x')\right) e^{iqx'}, 
\end{equation}
where $\mathcal{F}_q$ is the component of momentum $q$ for the function $V(x')\psi(x')$ after Fourier transformation. Then the general criterion for the vanished contribution of the extended term for Eq. \ref{eq-psi_x} requires that 
\begin{equation}
\mathcal{F}_{\pm p_{1}}\left(V(x')B_{2} \psi(x')\right)=0, 
\label{eq-BFqcomponent}
\end{equation}
where $\text{rank}(B_2) \le 1$. This condition can be understood intuitively that when $x$ is large enough when the localized component vanishes, only the extended component is important; yet this component vanishes. This condition can be used to determine the pole positions and hence the corresponding binding energy, which needs to lie between the two bands to satisfy the assumption of BIC, similar to Eq. \ref{eq-boundenergyB}. 

With this idea, we next consider a second case for the spin-orbit coupled model without any symmetry in Ref. \cite{kartashov2017boundstates},  with 
$A = \gamma \sigma_y p+\mu\sigma_{z}$, which can also support exact BIC. Using the same approach,  we have  
\begin{eqnarray}
    \psi(x)&=&\int_{-\infty}^{\infty}C_{1}(x-x') \exp(-\vert x-x'\vert\vert p_{2}\vert)V(x')B\psi(x')dx' \nonumber \\
    &+&\int_{-\infty}^{\infty}C_{2}(x-x') \cos{(( x-x') p_{1})} V(x')B\psi(x')dx' \nonumber \\
    &+&\int_{-\infty}^{\infty}C_{3} \sin{(\vert x-x'\vert p_{1})} V(x')B\psi(x')dx',  
    \label{eq-psi_x_2}
\end{eqnarray}
where
\begin{eqnarray}
    \frac{C_{1}}{2m^{2}} &=&\frac{E-\frac{p_{2}^{2}}{2m}}{\vert p_{2} \vert (p_{2}^{2}-p_{1}^{2})}+\mu\sigma_{z}+\frac{i\sigma_{y}\gamma \text{sign}(x-x')}{p_{2}^{2}-p_{1}^{2}},   \\ 
    \frac{C_{2}}{2m^{2}} &=& \frac{i\gamma\sigma_{y}}{p_{1}^{2}-p_{2}^{2}}\text{sign}(x-x'),    \\ 
    \frac{C_{3}}{2m^{2}} &=& -\frac{E-\frac{p_{1}^{2}}{2m}}{p_{1}(p_{1}^{2}-p_{2}^{2})}-\mu\sigma_{z},  
\end{eqnarray}
with $p_{1, 2}=\sqrt{2m}\sqrt{E+\gamma^{2}m\pm \sqrt{2Em\gamma^{2}+m^{2}\gamma^{4}+\mu^{2}}}$, assuming $p_{1}^{2}<p_{2}^{2}$. In general, this is also a quasi-BIC. To convert this quasi-BIC into an exact BIC, following Eq. \ref{eq-BFqcomponent}, the following new conditions should be satisfied
\begin{equation}
\mathcal{F}_{\pm p_{1}}(V(x')B\psi(x'))=0.
\label{eq-BFqcomponent2}
\end{equation}
This condition can simultaneously eliminate the cosine and sine terms with given momentum $\pm p_1$ in Eq. \ref{eq-psi_x_2}, yielding only a localized mode. The interpretation is the same as Eq. \ref{eq-BFqcomponent} for exact BIC. Furthermore, this condition can be used to determine $p_1$ and the binding energy. For this reason, these conditions for exact BIC play the same role as the continuum condition in Eq. \ref{eq-continuumcondition}. Hitherto, we can answer the fundamental question raised at the beginning of this work, that what is the fundamental reason to convert the quasi-BIC into an exact BIC. 

We verify the above predictions using the spin-orbit coupled model in Ref. \cite{kartashov2017boundstates} with the following potential 
\begin{equation}
    V(x)=\frac{2 \nu^{2}\left[3 \nu^{2}- \alpha \cosh ^{2}(\nu x)\right]}{\cosh ^{2}(\nu x)\left[ \alpha' \cosh ^{2}(\nu x)-\nu^{2}\right]}, 
\end{equation}
where $\alpha = \gamma^{2}+2 \nu^{2}+\sqrt{1-\gamma^{2} \nu^{2}}$, $\alpha' = 1+\sqrt{1-\gamma^{2} \nu^{2}}$, $\nu$ is a tuning parameter, $B = (\sigma_z + 1)/2$ and $E_{\text{BIC}}=-\nu^{2}/2+\sqrt{\mu^{2}-\nu^{2}\gamma^{2}}$. 
The wave function of the BIC (without the long tail) is presented in 
Fig. \ref{fig-fig3}(a). Meanwhile,  we have numerically verified that the constraint of Eq. \ref{eq-BFqcomponent2},  which is satisfied with $q=\pm p_1 = \pm 2.06325$. The results in case (IV) also mean that there is a lot of potential, not limited to the above ones, in which, by tuning the potential, we can convert the quasi-BIC into an exact BIC. This adjustability does not rely on symmetry \cite{Parker1966Resonance, Parker1967Resonance, Cobelli2009experimental, Moiseyev2009Suppression, Cederbaum2003Conical, Plotnik2011Experimental, Dreisow2009Adiabatic,Ni2016Tunable} or topology \cite{zhen2014topological, yang2013topological,Helical2015Vladimir}, and thus can be much more general and easily realized in experiments. The reason is that the exact BIC is a solution of the nonlinear equation, instead of the other properties. 

{\it Discussion and Conclusion}: The above analysis can be applied to multiband models immediately. Let us consider the following Hamiltonian
\begin{equation}
    H = {p^2 \over 2m}  + A + \mathcal{U}(x), 
\end{equation}
where $A$ is the coupling between the $N$ channels \cite{Friedrich1985interfering},  and the simplest choice of $\mathcal{U}(x)$ may be $\mathcal{U}(x)=\operatorname{diag}\{ V_1(x),  V_2(x),  \dots \}$. When $A = 0$,  this model is reduced to $N$ independent bands, which naturally host bound states in some channels. 
When the contribution of Fourier components of $\mathcal{U}(x') \psi(x')$ vanish at the real poles $p_i \in \mathbb{R}$, determined by $\text{det}(E - {p^2 \over 2m}  - A) =0$, see Eq. \ref{eq-BFqcomponent} and Eq. \ref{eq-BFqcomponent2}, then the quasi-BIC can be convertted into an exact BIC. Obviously, these poles can be controlled by $A$. With the increase of real poles, much more stringent tunability conditions are required to completely eliminate the extended components for an exact BIC using the coupling matrix $A$.

Our results will lead to important applications in different fields. Firstly, this state can be immediately realized using ultracold atoms, in which the coupling and the confining potentials can be tuned in experiments \cite{tunable2015Jim, luo2016tunable, Tunable2012Struck, zhang2013tunable}. Secondly, it can provide a new strategy to realize BIC in photonic systems using two coupled chains with tailored interaction between them  \cite{kang2023applications, Lee2012observation, Marinica2008Bound, koshelev2023bound, zhou2023increasing, minkov2019doubly, Yang2014Analytical, Vaidya2021Point, Optically2014Boretz,Bulgakov2014Bloch,yoon2015critical,Ni2016Tunable}. Thirdly, perhaps much more intriguingly, it can be generalized to many-body interactions in the Bose-Hubbard model  \cite{Huang2024Interaction, Yu2025RabiOscillation}, with nonlinear Kerr interaction in photonic lattice  \cite{Shit2025Intensity}, and gap and embedded solitons in the nonlinear Schrodinger equation \cite{zhang2013tunable, Yang2003Stable, Fan2020Gap}. We uncover its origin based on a bound state interacting with a continuum band, which can convert the quasi-BIC to an exact BIC by tuning their coupling, which is different from the previous literature based on symmetry and topology \cite{Parker1966Resonance, Parker1967Resonance, Cobelli2009experimental, Moiseyev2009Suppression, Cederbaum2003Conical, Plotnik2011Experimental, Dreisow2009Adiabatic,Ni2016Tunable, zhen2014topological, yang2013topological,Helical2015Vladimir}, and present the general criterion to convert a quasi-BIC into an exact BIC. It is hoped that this idea can provide a general approach to realize various BICs, which can greatly enrich our understanding of why bound states can coexist with the continuum spectra. 

\textit{Acknowledgments}: We thank Prof. Dezhuan Han for valuable discussion. This work is supported by the Strategic Priority Research Program of the Chinese Academy of Sciences (Grant No. XDB0500000), National Key Research and Development Project (Grant No. 2024YFA1410900, No. 2023YFB3610500), National Natural Science Foundation of China (No. U23A2074, No. 62275241, No. 92265210 and No. 12293053), and the Innovation Program for Quantum Science and Technology (2021ZD0301200, 2021ZD0301500). 

\bibliography{ref.bib}

\begin{thebibliography}{86}
\bibitem [{bic()}]{bicperiodic}%
  \BibitemOpen
  \href@noop {} {}\bibinfo {note} {This is different from the BIC in periodic system, in which the square integrable of the Bloch wave function is only performed in one unit cell, in which the linewidth can still approach zero, with infinite long lifetime \cite{zhen2014topological}.}\BibitemShut {Stop}%
\end{thebibliography}

\end{document}